\documentclass[aps,10pt,twocolumn,groupedaddress,pre,showpacs]{revtex4-1}
\usepackage{colordvi,epsfig,color,amsmath,amssymb}

\begin{document}

\def\be{\begin{equation}}
\def\ee{\end{equation}}
\def\bee{\begin{eqnarray}}
\def\eee{\end{eqnarray}}
\def\sech{\mbox{sech}}
\def\e{{\rm e}}
\def\d{{\rm d}}
\def\L{{\cal L}}
\def\U{{\cal U}}
\def\M{{\cal M}}
\def\T{{\cal T}}
\def\V{{\cal V}}
\def\R{{\cal R}}
\def\kb{k_{\rm B}}
\def\tw{t_{\rm w}}
\def\ts{t_{\rm s}}
\def\Tc{T_{\rm c}}
\def\gs{\gamma_{\rm s}}
\def\tm{tunneling model }
\def\TM{tunneling model }
\def\tilde{\widetilde}
\def\Deltac{\Delta_{0\rm c}}
\def\Deltamin{\Delta_{0\rm min}}
\def\Emin{E_{\rm min}}
\def\tauc{\tau_{\rm c}}
\def\tauac{\tau_{\rm AC}}
\def\tauw{\tau_{\rm w}}
\def\taumin{\tau_{\rm min}}
\def\taumax{\tau_{\rm max}}
\def\de{\delta\varepsilon / \varepsilon}
\def\pF{{\bf pF}}
\def\pFAC{{\bf pF}_{\rm AC}}
\def\halb{\mbox{$\frac{1}{2}$}}
\def\dreihalb{\mbox{$\frac{3}{2}$}}
\def\viertel{\mbox{$\frac{1}{4}$}}
\def\achtel{\mbox{$\frac{1}{8}$}}
\def\with{\quad\mbox{with}\quad}
\def\und{\quad\mbox{and}\quad}
\def\za{\sigma_z^{(1)}}
\def\zb{\sigma_z^{(2)}}
\def\ya{\sigma_y^{(1)}}
\def\yb{\sigma_y^{(2)}}
\def\xa{\sigma_x^{(1)}}
\def\xb{\sigma_x^{(2)}}
\def\spur#1{\mbox{Tr}\left\{ #1\right\}}
\def\erwart#1{\left\langle #1 \right\rangle}
\def\tr{{\rm tr}}
\newcommand{\bbbone}{{\mathchoice {\rm 1\mskip -4mu l}{\rm 1\mskip 
-4mu l}{\rm 1\mskip -4.5mu l}{\rm 1\mskip -5mu l}}}

\newcommand{\bj}{{\bf j}}
\newcommand{\bx}{{\bf x}}

\title{Exciton dynamics and Quantumness of energy transfer in the Fenna-Matthews-Olson complex}

\author{P. Nalbach$^1$, D.\ Braun$^2$, and M. Thorwart$^1$}
\affiliation{$^1$I.\ Institut f\"ur Theoretische Physik,  Universit\"at
Hamburg, Jungiusstra{\ss}e 9, 20355 Hamburg, Germany \\
$^2$ Laboratoire de Physique Th\'{e}orique, Universit\'{e} Paul Sabatier, 118,
route de Narbonne, 31062 Toulouse, France}

\date{\today}

\begin{abstract}
We present numerically exact results for the quantum coherent energy 
transfer in  the Fenna-Matthews-Olson molecular aggregate under 
realistic physiological conditions, including vibrational fluctuations 
of the protein and the pigments for an experimentally determined fluctuation 
spectrum. We find coherence times shorter than observed experimentally.
Furthermore we determine the energy transfer current and quantify its 
``quantumness'' as the distance of the density matrix to the classical 
pointer states for the energy current operator.  
Most importantly, we find that the energy transfer happens through 
a ``Schr\"odinger-cat'' like superposition of energy current pointer 
states.  
\end{abstract}
\pacs{03.65.Yz, 03.65.Ud, 05.30.-d, 87.15.H-}
\maketitle

\section{Introduction}

Recent experiments on the ultrafast exciton dynamics in photosynthetic
biomolecules
have brought a longstanding question again into the scientific focus whether 
nontrivial quantum coherence effects exist in natural biological systems under 
physiological conditions, and, if so, whether they have any functional significance.
Photosyn\-thesis \cite{vanAmerongen00} 
starts with the harvest of a photon by a pig\-ment and the 
formation of an exciton, followed by its transfer to the
reaction center, where charge separation via a primary electron transfer is initiated. The
transfer of excitations has traditionally been regarded as an
in\-coherent 
hopping between molecular sites \cite{MayKuehn}. 

Recently, Engel et al.~\cite{EngelFleming07,Engel10} have reported longlasting beating signals
in time-resolved optical two-dimensional spectra of the Fenna-Matthews-Olson
(FMO)~\cite{Fenna75,Milder10} 
complex, which have been interpreted as 
evidence for quantum coherent energy transfer via delocalized exciton states. 
It transfers the excitonic energy from the chlorosome to the reaction
center and consists  of three identical subunits, each with seven bacteriochlorophyll molecular
sites (the existence of an eighth site is presently under investigation). 
Quantum coherence times of more than 660 fs
 at $77$~K \cite{EngelFleming07} and about $300$ fs at physiological temperatures
\cite{Engel10} have been reported.  

Together with recent experiments \cite{Scholes10} on marine cryptophyte algae, 
these reports have boosted on-going research to answer the question how quantum 
coherence can prevail over such long times in a strongly fluctuating physiological 
environment of strong vibronic protein modes and the surrounding polar solvent.
Theoretical modeling of the real-time dynamics is notoriously complicated due to the
large cluster size and  strong non-Markovian fluctuations. It relies
on simple models \cite{Reineker82} of few 
chromophore sites which interact by dipolar couplings and which are exposed to
fluctuations of the solvent and the protein  
\cite{vanAmerongen00,MayKuehn}. A calculation of the 2D optical spectrum assuming
weak coupling to the environment does not reproduce the experimental data \cite{Sharp10}.
It also became clear that standard Redfield-type approaches fail even for 
dimers~\cite{Ishizaki09a,Nalbach10a}.

In connection with molecular exciton transfer, extensions beyond perturbative
approaches, such as
the stochastic Liouville equation \cite{Reineker82}, extended Lindblad approaches
\cite{Mukamel2009} or small polaron approaches \cite{Jang08}, have been
formulated.
Two recent approaches have included parts of the standard FMO model. 
A second-order cumulant time-nonlocal quantum master equation found
coherence times \cite{AkiPNAS09} as observed experimentally for the 
FMO complex. However, they employed an Ohmic bath in which the
relaxation time is not uniquely fixed by experiments and excluding 
strong vibrational protein modes. 
On the other hand, a variant of a time-dependent 
DMRG scheme \cite{Plenio10} has been applied to a dimer in a
realistic FMO environment \cite{Adolphs06}. 
The solution of the full state-of-the-art FMO model with seven localized 
sites and a physical environmental spectrum \cite{Adolphs06} including a
strongly localized Huang-Rhys mode \cite{MayKuehn} is still missing. 
In particular, it has not been clarified whether a strong vibronic mode still
allows for the observed quantum coherence times.  

In this paper we close this gap and present numerically exact results for 
the full FMO model with seven sites and the spectral density of Ref.\ \cite{Adolphs06},
determined from optical spectra \cite{Pullerits00}. 
Upon adopting the iterative real-time  quasiadiabatic propagator path-integral (QUAPI)
\cite{Makri95a,Thorwart98} scheme we find coherence times shorter than
observed experimentally. 
For comparisson we recalculate also the dynamics employing the same
Ohmic fluctuation spectrum as Ishizaki and Fleming \cite{AkiPNAS09}.
To quantify quantum effects for energy transfer, 
we calculate the energy current
associated with the transfer dynamics and its ``quantumness''. It has 
been argued that quantum entanglement could be created {\em in a single excitation
sub-space\/} during the energy transfer
\cite{Whaley}, but this form of entanglement 
cannot be used to violate a Bell-inequality \cite{Beenakker06} and its role
for the transfer efficiency is therefore unclear. Wilde et al. \cite{Wilde}
used the Leggett-Garg inequality to discuss whether quantum effects are relevant
but they applied a phenomenological Lindblad approach.
Here we use a recently
developed measure of quantumness based on the Hilbert-Schmidt distance of
the density matrix to the convex hull of classical states
\cite{Giraud10}, taken as the ``pointer-states''
\cite{Zurek81} of the energy-current operator.  
We show that energy transfer starts with large current out of the initial
site and then small currents flow between all seven sites. All currents
show substantial quantumness.
Interestingly enough, this implies that the energy
transfer happens through a largely coherent ``Schr\"odinger-cat'' like
superposition of pointer states of the energy current operator. 

In the next section we shortly recapitulate the Hamiltonian and the
fluctuation spectrum of the site energies for the Fenna-Matthews-Olson
complex. Then we determine the population dynamics within the single 
excitation subspace. In the fourth section we calculate the energy currents
between any two sites in the FMO and then introduce a measure for the
quantumness of these currents. Finally we conclude with a short summary of 
our results.

\section{FMO Model}

The FMO complex is a trimer consisting of identical, weakly
interacting monomers \cite{Adolphs06}, each containing seven
bacteriochlorophyll{\it a} (BChl{\it a}) molecular sites which transfer
excitons. The pigments are embedded in a
large protein complex. Each of it can be
reduced to its two lowest electronic levels and their excited states are
electronically coupled along the
complex. Recombination is negligibly slow ($\sim$ ns) compared to exciton transfer times 
($\sim$ps). Thus, the excitation dynamics is
reliably described within the 1-exciton subspace. The coupling of the seven 
excited levels gives rise to the Hamiltonian 

\be \label{hfmo}
 H_{\rm FMO} = \\
 {\footnotesize \left( 
\begin{array}{ccccccc}
240 & -87.7 &   5.5 &   -5.9 &   6.7 & -13.7 &  -9.9 \\
    & 315   &  30.8 &    8.2 &   0.7 &  11.8 &   4.3 \\ 
    &       &   0   &  -53.5 &  -2.2 &  -9.6 &   6.0 \\
    &       &       &  130   & -70.7 & -17.0 & -63.3 \\
    &       &       &        & 285   &  81.1 &  -1.3 \\
    &       &       &        &       & 435   &  39.7 \\
    &       &       &        &       &       & 245 
\end{array}\right) } 
\ee
in units of cm$^{-1}$ in site representation \cite{Adolphs06} 
for an FMO monomer of 
{\it C.\ tepidum}. We define the lowest site energy of pigment $3$ as
reference. 

The vibrations of the BChl{\it a}, the embedding protein and the surrounding polar solvent are too complex
for a pure microscopic description and are thus treated with the framework of
open quantum systems. They induce thermal fluctuations described by harmonic
modes \cite{MayKuehn} and lead to the total
Hamiltonian \cite{Adolphs06}
\bee
\label{Htot} H &=& H_{\rm FMO} \\
&& + \sum_{j=1}^7|j\rangle\langle j|
\sum_{\kappa}\nu_{\kappa}^{(j)} q_{j,{\kappa}} \,+ \sum_{j=1}^7 \halb\sum_{\kappa}\left(
p_{j,{\kappa}}^2+\omega_{j,{\kappa}}^2q_{j,{\kappa}}^2\right)  \nonumber
\eee
with momenta $p_{j,{\kappa}}$, displacement $q_{j,{\kappa}}$, frequency $\omega_{j,{\kappa}}$ and coupling
$\nu_{\kappa}^{(j)}$ of the environmental vibrations at site $j$. 
We assume that fluctuations at different sites are identical but spatially uncorrelated
\cite{Kleinekathoefer10c}. 

\begin{figure}[t!]
\epsfig{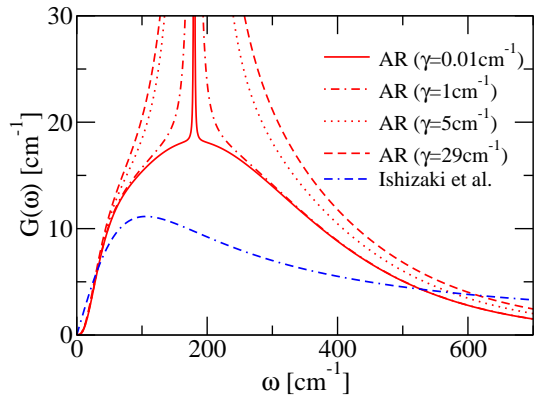}
\caption{\label{fig1} (Color online) Bath spectral densities of the
FMO complex of {\em C. tepidum}: The most realistic form as determined up to
present \cite{Adolphs06,
Pullerits00} is given by Eq.\ (\ref{fullspecdens}) (AR, red (upper four) lines). The blue
dash-dot-dashed (lower) line shows the Drude form $G_D(\omega)$ used in \cite{AkiPNAS09}.} 
\end{figure}

The key quantity which determines the FMO coherence properties is the environmental spectral density 
$G(\omega)=\sum_{j,\kappa}(|\nu_{\kappa}^{(j)}|^2/2\omega_{j,{\kappa}}) \delta(\omega-\omega_{j,{\kappa}})$.  
The most detailed identification of the bath properties up to date has been
achieved by Adolphs and Renger \cite{Adolphs06}. They used an advanced theory of
optical spectra, a genetic algorithm, and
excitonic couplings from electrostatic calculations modeling the dielectric
protein environment to derive the up to present most detailed spectral density  
\be
\label{fullspecdens}
 G_{AR}(\omega) \,=\, \omega^2 S_0 g_0(\omega) \,+\omega^2 S_H \delta(\omega-\omega_H) \, .
\ee
Here, $S_0=0.5$, $S_H=0.22$, $\omega_H=180$cm$^{-1}$ and
\[ g_0(\omega) = 6.105\cdot10^{-5}\cdot \frac{\omega^3}{\omega_1^4} e^{-\sqrt{\frac{\omega}{\omega_1}}}
+ 3.8156\cdot10^{-5}\cdot \frac{\omega^3}{\omega_2^4} e^{-\sqrt{\frac{\omega}{\omega_2}}}
\]
with $\omega_1=0.575$ cm$^{-1}$ and $\omega_2=2$ cm$^{-1}$. It includes a broad continuous
part (see red (upper four) lines in Fig.\ \ref{fig1}) which for $\omega \to 0$ 
behaves as super-Ohmic,  $G_{AR}(\omega) \sim \omega^5$, and
which describes the protein vibrations with the Huang-Rhys factor $S_0$. It was
determined from temperature-dependent absorption spectra
\cite{Pullerits00}. In addition, 
 a vibrational mode of the individual pigments with the
Huang-Rhys factor $S_H$ is included.
We have added a broadening to the unphysical $\delta$-peak,
which is justified since the protein is embedded in water as a
polar solvent which gives rise to an additional weak Ohmic damping of the
protein vibrations. We fix its width to 
$\gamma_p=29$ cm$^{-1}$ which was found for the lowest energy peak of protein
vibrations in the LH2 complex \cite{Kleinekathoefer10}.

In order to recover known results for the transfer dynamics of excitations 
in the Fenna-Matthews-Olson (FMO) complex embedded in an Ohmic bath \cite{AkiPNAS09}, 
we additionally consider the spectral density used in Ref.\ \cite{AkiPNAS09} given by 
\be
\label{debye}
G_D(\omega) = \frac{2\lambda \omega \omega_c}{\pi(\omega_c^2+\omega^2)}
\ee
with a Debye cut-off at frequency $\omega_c$ (see blue dash-dot-dashed (lowest) line in Fig.\ \ref{fig1}). 
It includes the 
reorganization energy $\lambda=35$ cm$^{-1}$ and the environmental 
timescale 
$\omega_c^{-1}=50$ fs.


\section{Population dynamics in the FMO}

\begin{figure}[t!]
\epsfig{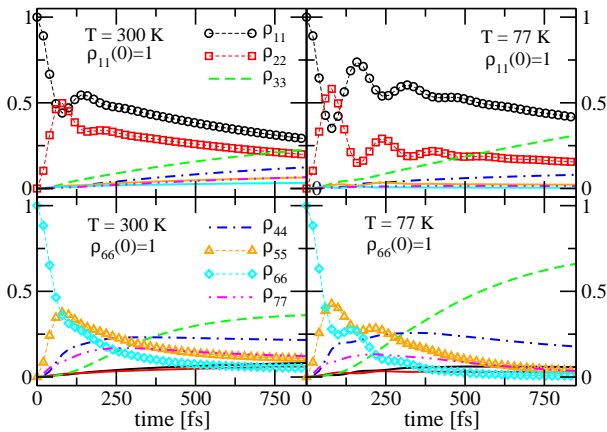}
\caption{\label{fig2} (Color online) Time-dependent occupation probabilities of
all seven FMO sites for $T=300$ K, $77$ K with $\rho_{11}(0)=1$ and
$\rho_{6}(0)=1$ for the measured FMO spectrum Eq.\
(\ref{fullspecdens}) \cite{Adolphs06}. }
\end{figure}

To understand whether the vibrational fluctuations of the protein
and the pigments allow for the observed coherence times, 
we apply the numerically exact quasi-adiabatic propagator
path-integral (QUAPI) to
simulate the real time exciton dynamics in the FMO complex for the
realistic bath spectral density (\ref{fullspecdens}). QUAPI is well established
\cite{Makri95a,Thorwart98} and allows
us to treat nearly arbitrary spectral functions at finite temperatures. We 
extended recently the original scheme to treat multiple environments, i.e.
separate environments for each chromophore site \cite{Nalbach10b}.

We calculate the time-dependent populations of the FMO pigment sites 
for the spectral density of Eq.\ (\ref{fullspecdens}).
We choose $T=300$K (physiological temperature) and $T=77$K
(typical experimental temperature). Both the pigments BChl $1$ and
BChl $6$ are oriented towards the baseplate protein and are thus believed to be
initially excited (entrance sites) \cite{Wen09}. Thus, we consider 
two cases, i.e., $\rho_{11}(0)=1$ and $\rho_{66}(0)=1$. 
We focus on the transient coherence effects and thus do not
include an additional sink at the exit site BChl $3$. 
In Fig.\ \ref{fig2}, we show the time-dependent pigment occupation probabilities
$\rho_{jj}(t)$.
Identical simulations using smaller width, i.e. $\gamma_p=5$ cm$^{-1}$ and 
$\gamma_p=1$ cm$^{-1}$, for the vibrational mode yield identical
results for the populations (not shown).
For $\rho_{66}(0)=1$ at room temperature,
coherent oscillations are completely suppressed
and at $77$ K they last up to about $250$ fs. For $\rho_{11}(0)=1$, coherence is
supported longer due to the strong electronic coupling
between sites $1$ and $2$. At room temperature, it survives for up
to about $200$ fs and for $77$ K up to $500$ fs at most. 
Thus, coherence times are shorter than experimentally observed. 
We emphasize that coherence features are a transient property and, hence, not
only the low frequency bath modes, but also the discrete vibrational modes are
relevant ($100$ fs corresponds to $\sim 333$ cm$^{-1}$).

We additionally calculate the time-dependent populations of the FMO pigment sites for 
$T=300$K (physiological temperature) and $T=77$K
(typical experimental temperature) for the FMO with the bath spectrum $G_D(\omega)$
(see Eq.\ (\ref{debye})). As above we consider 
 two choices of initial conditions, i.e., $\rho_{11}(0)=1$ 
and $\rho_{66}(0)=1$ and focus on the transient coherence effects and thus do not
include an additional sink at the exit site BChl $3$. 

\begin{figure}[t!]
\epsfig{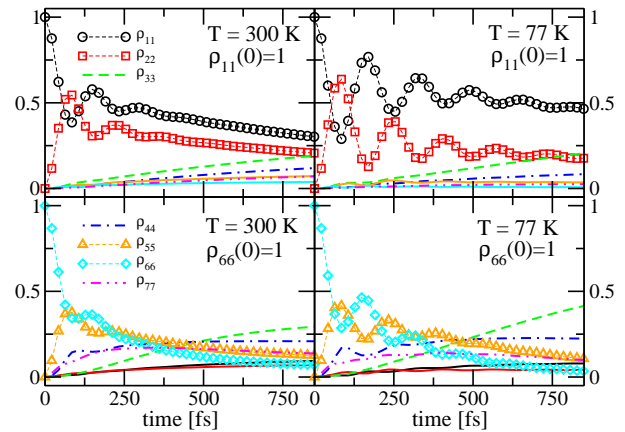}
\caption{\label{fig2b} (Color online) Time-dependent occupations of the seven FMO
sites for $T=300$ K and $T=77$ K with $\rho_{11}(0)=1$ and $\rho_{66}(0)=1$ for
the Ohmic spectrum $G_D(\omega)$.}
\end{figure}

Fig. \ref{fig2b} shows the occupation probabilities of all seven
sites at $T=300$ K versus time. 
For both initial conditions, coherent oscillations
of the populations of the initial site
and its neighboring site ($j=2$ for $\rho_{11}(0)=1$ and $j=5$ for
$\rho_{66}(0)=1$) occur due to the strong electronic couplings. They 
last up to $\sim 350$ fs.  
At $T=77$ K, as shown in Fig. \ref{fig2b} (b) and \ref{fig2b}(d), the 
oscillations persist up to $\sim 700$ fs. Our results coincide with those of 
Ref.\ \cite{AkiPNAS09}. Tiny deviations at short times arise since 
the full FMO trimer is considered in Ref.\ \cite{AkiPNAS09}. The
coherence times, however, are not affected. 

\section{Quantumness of Energy transfer through FMO}

Next, we discuss whether energy transfer dynamics is 
a quantum coherent process. Classical coupled dipoles can show oscillatory 
energy currents identical to quantum mechanical dipole-coupled systems provided that
the interaction between the dipoles is weak enough to be treated via 
linear response \cite{Zimanyi}. In FMO (see Eq.\ \ref{hfmo}), however, off-diagonal
couplings are on the same order of magnitude as site energy differences. 
We use a physically well motivated measure of quantumness which can be found by first identifying 
the most classical pure states in the problem. In quantum optics, for 
instances, coherent states are considered ``(most) classical'' due to
their minimal uncertainty and their stability under the free evolution
of the electro-magnetic field \cite{Scully97}. Next, one identifies
the set of all classical states with the convex hull (i.e.~all 
mixtures) of the pure classical states, as mixing cannot
increase quantumness.  
This leads in quantum optics and for spin-systems to the
well established notion that classical states are those with a
well-defined positive 
$P$-function \cite{Mandel,Knight,Giraud08,Giraud10}.  
Finally, one defines ``quantumness'' as the 
distances to the closest classical states. 

For FMO we are interested in the ``quantumness'' of the energy
current since energy transfer constitutes the main function of the 
FMO complex. A measurement of the energy current would yield that 
for any measurement result the state of the system would collapse onto
a pointer state. In general, pointer states
are the states einselected by the interaction with
its environment \cite{Zurek81} being here the hypothetical measurement
apparatus for measuring energy current.  Pointer states are the most
classical states in that they arise via decoherence and are those that 
persist when a quantum system interacts strongly and for a long time 
with its environment.
In our case of a projective von Neumann measurement, they
are the eigenstates of the quantum mechanical operator
corresponding to the energy current measurement.

\subsection{Energy currents in FMO}

An energy current operator can be derived
for a general multi-site Hamiltonian \cite{Hardy63}.
The expression for an energy current operator $\bj(\bx)$ can be obtained
from a
continuity equation 
\begin{equation} \label{cont}
\dot{H}(\bx)+\nabla\cdot\bj(\bx)=0\,,
\end{equation}
where $H(\bx)$ is the local energy density. This is a well-established
procedure introduced by Hardy \cite{Hardy63}, and which was
largely adopted by a large number of
subsequent papers
(see e.g.~\cite{deppe_energy_1994,segal_thermal_2003,allen_thermal_1993,leitner_frequency-resolved_2009,wu_energy_2009}.)
There is some freedom in the definition of local energy density, but for a
Hamiltonian of the  
form $H=\sum_i T_i+(1/2)\sum_{i\ne k}V_{ik}$, where $T_i$ and $V_i$ are
kinetic 
energy and interaction energy (of possibly a large number of particles) in
region $i$, and $V_{ik}$ 
the 
interaction between the particles in regions $i$  
and those in region $k$, a natural decomposition is
given by $H=\sum_i h_i$, $h_i=T_i+(1/2)\sum_{k\ne i}V_{ik}\equiv v_iH(x_i)$,
where $v_i$ is a volume of region $i$. \\

For the FMO complex, precise information about the
the single-excitation 
sector of the macro-molecule with sites $i\in \{1,\ldots,N\}$ is known
($N=7$). The Hamiltonian $H= H_{FMO}$ is given in tight-binding approximation as
\begin{equation} \label{H}
H=\sum_{i,k}^Nh_{ik}|i\rangle\langle k|\,,
\end{equation}
where $|i\rangle$ is a state with the excitation localized on site $i$ (see
Eq.(1) ).  The
natural decomposition of that Hamiltonian in terms of local excitations is 
\begin{equation} \label{Hloc}
H=\sum_{i=1}^N h_i\mbox{ with } h_i=\frac{1}{2}\sum_{k=1}^N \left(h_{ik}|i\rangle\langle k|+h.c.\right)\,.
\end{equation}
In order to deduce an energy-current operator from this Hamiltonian,
consider first a 
linear chain in 1D, where $\bj$ has only one component, denoted by $j$, with
values $j_i$ on sites $i$, taken as equidistant with lattice constant $a$.
The discretized form of (\ref{cont}) reads 
\begin{equation} \label{contd}
\frac{\partial}{\partial t} \frac{h_i}{a}=\frac{j_i^{l}-j_i^{r}}{a}\,,
\end{equation}
where $j_i^{l,r}$ denote the energy flux in positive $x$-direction on the
left of site $i$ and on the right of site $i$. It is convenient to think of
$j_i^{l}$ ($j_i^{r}$) as being evaluated half way between sites $i-1$ and
$i$ (between sites $i$ and $i+1$), respectively.  Current $j_i^{l}$ arises
from the balance of the currents $s_{i-1\to i}$ from site $i-1$ to site $i$
and $s_{i\to i-1}$ from site $i$ to site $i-1$.  The latter two currents are
always defined positive and in the direction indicated by the indices,
whereas $j_i$ can be positive or negative, depending on which current
component dominates.  Inserting these expressions into (\ref{contd}), we
find 
\begin{equation} \label{cont1D}
\frac{\partial }{\partial t} h_i=\sum_{k=i\pm 1}(s_{k\to i}-s_{i\to k})\,.
\end{equation}
This equation is valid for the 1D tight-binding model with only nearest
neighbor couplings. It has the natural interpretation that the change of
energy density at a given site is given by the difference between the sum of
incoming energy currents and the sum of outgoing energy currents. As such, the
expression generalizes in straight-forward fashion to a general
tight-binding model on a graph, where each site can be connected to an
arbitrary number of other sites.  Each link can support an energy current, and
one has thus
\begin{equation} \label{contG}
\frac{\partial }{\partial t} h_i=\sum_{k\ne i}(s_{k\to i}-s_{i\to k})\,.
\end{equation}
The left-hand side of (\ref{contG}) is easily calculated by using
Heisenberg's equation of motion,
\begin{equation} \label{heisi}
\frac{\partial }{\partial t}h_i=\frac{i}{\hbar}[H,h_i]\,.
\end{equation}
A short calculation leads to 
\begin{equation} \label{htbG}
\frac{\partial }{\partial t}h_i=\frac{i}{2\hbar}\sum_{k\ne
  i}\sum_l\left(h_{ki}h_{il}|k\rangle\langle l|-h_{ik}h_{kl}|i\rangle\langle
  l|-h.c.\right)\,. 
\end{equation}
Comparing this expression with (\ref{cont1D}), we identify the directed
energy current 
\begin{equation} \label{sik}
s_{i\to k}=\frac{i}{2\hbar}\sum_l\left(h_{ik}h_{kl}|i\rangle\langle
l|-h.c.\right)\,. 
\end{equation}
The (positive or negative) energy current attributed to link
$i-k$ on the graph is $j_{ik}=s_{i\to k}-s_{k\to i}$. The dimension of all
these energy currents is energy/time.  A
corresponding classical energy current phase space function can be derived
in classical mechanics, if a corresponding classical 
Hamilton function can be established, in a completely analogous way by
using Hamilton's 
equation of motion to derive a continuity equation for energy transport. 

\begin{figure}[t!]
\epsfig{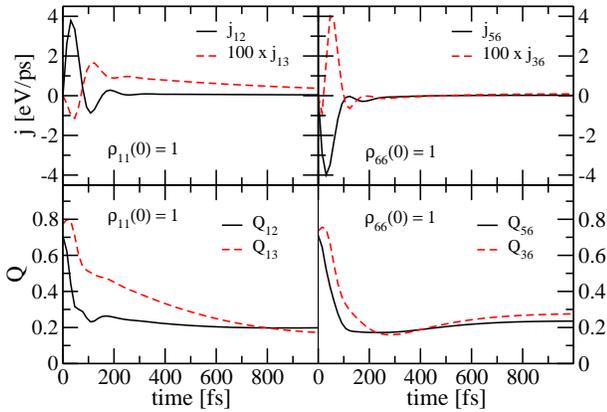}
\caption{\label{figJQ} (Color online) Selected average energy currents $j_{ik}$
  (top panels)
  and  corresponding
  quantumness $Q_{ik}$ (lower panels) for FMO spectral density of Eq.\
(\ref{fullspecdens}) and $T=300$K. Notice the rescaling of the currents 
 for better visibility. }
\end{figure}

\begin{figure}[t!]
\epsfig{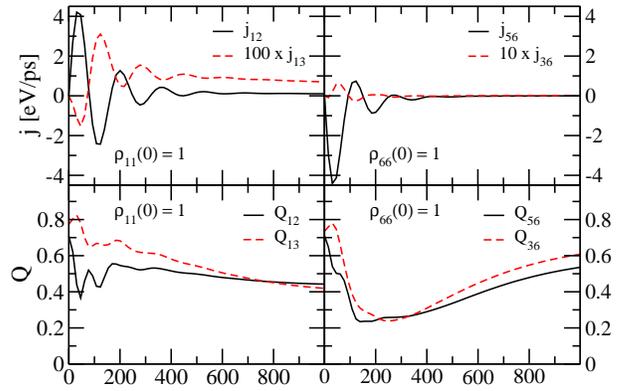}
\caption{\label{figJQB} (Color online) Same as Fig.\ref{figJQ}, but at
$T=77$K.}
\end{figure}

We have calculated all energy currents $\langle
j_{ik}(t)\rangle=\tr[j_{ik}\rho(t)]$ for all sites $i$ and $k$.
The currents out of the entrance site towards its strongest coupled
neighbor ($j_{12}$ for
$\rho_{11}(0)=1$ and $j_{56}$ for $\rho_{66}(0)=1$) have by far the largest
amplitude. 
All currents show initial oscillations which
are more pronounced at 77K than at 300K (see
Fig.~\ref{figJQ} and Fig.\ref{figJQB}) on times scales comparable to coherent
oscillations discussed above. The currents out of the entrance site towards its 
strongest coupled {\it neighbor} ($j_{12}$ for $\rho_{11}(0)=1$ and $j_{56}$ 
for $\rho_{66}(0)=1$) have by far the largest amplitude. 
After this initial phase energy currents between all sites are small but finite
(as exemplary shown by $j_{13}$ and $j_{36}$) and persist up to 1000fs.
Energy transfer, thus, starts by quantum coherent population exchange mainly 
between site $1$ and $2$ ( or $6$ and $5$ respectively) and some transfer to 
other sites. After this initial phase,
in which a prelimenary redistribution of energy takes place, small 
currents between all sites will slowly bring the system into thermal equilibirum
and result in an according population / energy distribution.

\subsection{Classical states of energy transport}

It has long been appreciated that the most classical states corresponding to
a given quantum mechanical observable are the so-called ``pointer states''.
These states are ``einselected'' by the decoherence process due to the
interaction of an environment with the system.  In the case of a
measurement, the environment is given by the measurement instrument, and the
pointer states are in this case simply the eigenstates of the operator
representing the quantum mechanical observable \cite{Zurek81,Zurek03}. It is
in the eigenbasis of these pointer states that the density matrix becomes
diagonal due to the decay of the off-diagonal 
matrix elements.  The remaining diagonal matrix elements correspond to the 
probabilities of finding one of the possible outcomes of the measurement.\\

As eigenstates of a Hermitian operator, pointer states are pure states.
The most general classical states can be obtained by classically mixing
the pointer states, i.e., one chooses pointer states randomly with
a certain probability.  As this is a purely classical procedure, it cannot
increase the ``quantumness'' of the system. This definition of
``classicality'' is 
well-established in quantum optics and leads to a criterion of classicality
based on a well-defined positive-definite $P$-function
\cite{mandel86,Kim05}.  It was recently extended to spin-systems
\cite{Giraud08,Giraud10}. \\

\subsection{Quantumness}

Thus, if one wants to decide whether the energy current between two
different sites in FMO has
to be considered ``quantum'', one has to ({\em i}) find the pointer states
of the corresponding energy current operator, and ({\em ii}) check whether
the state of the system $\rho(t)$ can be written as a convex combination of
these 
pointer states.  A more quantitative statement is possible by measuring the
distance of $\rho(t)$ to the convex set of classical states
\cite{Giraud08,Giraud10,Martin10}, i.e., here the convex hull of the pointer
states. 
The pointer states are the eigenvectors $|v_{ik}^{l}\rangle$ of  $j_{ik}$. They define the
relevant
pure classical states for the energy transfer along link $i-k$. Their
classical mixtures form the 
convex set of all classical states for the energy current on link $i-k$. 
Any distance 
measure in Hilbert space is in principle suitable, and we use the
Hilbert-Schmidt distance for simplicity, leading to our definition of
quantumness of the current $j_{ik}$,  
\begin{equation} \label{Qik}
Q_{ik}(\rho)=\min_{\{p_i|p_i\ge 0,\sum_i p_i =1\}}||\rho-\sum_l
p_l |v_{ik}^{l}\rangle  \langle v_{ik}^{l}|||\,
\end{equation}
with $||A||=(\tr AA^\dagger)^{1/2}$.
Note that this measure of ``quantumness'' is completely analogous to and on
the 
same level of abstraction as the
measure of entanglement based on the distance of a state to the convex
set of separable states (see p.363 in Ref. \cite{Bengtsson06}), but has the
advantage of being meaningful even in the single-excitation sector, and
without artificially separating the system into two subsystems. \\
A finite value of $Q_{ik}$ means that the state cannot be written as a
convex sum of pointer-states of the energy current operator $j_{ik}$,
i.e., there are coherences left in $\rho(t)$ written in the pointer
basis. This means essentially, that the system is in a Schr\"odinger-like
cat state of different energy-current pointer states at a given time $t$.

By definition,
$Q(\rho)\ge 0$, and $Q(\rho)=0$ if $\rho$ is classical. An upper
bound is
given by $Q(\rho)\le Q_{\rm max} \equiv \sqrt{\tr \rho^2-1/d}$,
where $d$ is the dimension of the 
Hilbert space \cite{Giraud10} ($Q_{\rm max}\simeq 0.925$ for pure states and
$d=7$). 
We have calculated the quantumness $Q_{ik}(t)$ for all energy currents $\langle
j_{ik}(t)\rangle$. As discussed above the currents out of the entrance site 
towards its strongest coupled {\it neighbor} ($j_{12}$ for
$\rho_{11}(0)=1$ and $j_{56}$ for $\rho_{66}(0)=1$) have by far the largest
amplitude and all currents show initial oscillations which
are more pronounced at 77K than at 300K (see
Fig.~\ref{figJQ} and Fig.~\ref{figJQB}) on time scales comparable to coherent
oscillations discussed above. These oscillations go
hand-in-hand with substantial quantumness: $Q_{12}$ and $Q_{13}$ are
initially of the order of $Q\simeq 0.6-0.8$, but drop within $\sim 100$fs to
$Q\simeq 0.2-0.4$, with the exception of the case ($\rho_{11}(0)=1$, $T=77$K),
where $Q_{13}$ drops slowly over 1000fs, much more slowly than the oscillations
of the energy current, and the rapid decay of
coherences in the site basis not withstanding.
After this initial phase energy currents between all sites are small but finite
(as exemplary shown by $j_{13}$ and $j_{36}$) and persist up to 1000fs and,
noteworthy, a substantial amount of
quantumness of the order $Q\simeq 0.2-0.4$ still persists at 1000fs as well.
In the last graph ($\rho_{66}(0)=1$, $T=77$K), the quantumness even rises again
after the initial drop. 

This implies that the superposition of the pointer states of the
energy current operator remains largely coherent even after population
dynamics does not show coherence anymore. Nature has
apparently engineered the environment of the FMO complex in such a way
that it is rather inefficient in decohering superpositions of energy
currents. Thus even though the actual currents, which bring the system into
thermal equilibrium, are rather small (after the initial oscilatory phase), 
they turn out to have a finite {\it quantumness}.
This is further illustrated by comparing with the quantumness 
of the thermal equilibrium state $\rho_T=\exp(-H_{\rm FMO}/k_BT)$ at
$T=77$K and 
$T=300$K, which is of the order 0.01 
and 0.002, respectively.  
Altogether, our data quantitatively show the
non-classical nature of the energy transfer in FMO, and
provide a clear physical picture of the quantumness: energy
transport in FMO happens largely through a coherent
Schr\"odinger-cat like superposition of
pointer states of the energy current operator. \\

\section{Conclusion}

To summarize, we have obtained numerically exact results for the real-time
exciton dynamics of the FMO  complex in presence of realistic environmental
vibronic fluctuations. We have used the most accurate form of the 
spectral density realized in nature. It includes vibronic effects via a strongly
localized Huang-Rhys mode. The resulting coherence times of the populations 
are shorter than observed experimentally. The energy transfer dynamics is 
also intrinsically quantum mechanical on the time scales of
several hundred femtoseconds. This has been shown by calculating 
the quantumness as distance to the convex hull of pointer states of 
the energy current. The energy transport in FMO is to a large extent through 
a coherent Schr\"odinger-cat like superposition of the pointer states.

\section*{Acknowledgements}

We thank T.\ Pullerits and P.~A.~Braun for fruitful discussions.

\end{document}